\documentclass{PoS}
\usepackage{lineno}
\title{Studying Cosmic-ray Interactions in Giant Molecular Clouds with the HAWC Gamma-ray Observatory}

\ShortTitle{Giant Molecular Clouds with HAWC}

\author{
\speaker{Hugo Alberto Ayala Solares$^{1}$} \\
The HAWC Collaboration\footnote{For collaboration list, see PoS(ICRC2019) 1177.}\\
{\itshape \href{https://www.hawc-observatory.org/collaboration/icrc2019.php}{https://www.hawc-observatory.org/collaboration/icrc2019.php}}\\
{$^{1}$ \itshape The Pennsylvania State University}\\
E-mail: \email{hgayala@psu.edu}
}

\abstract{
The cosmic-ray flux in the Galaxy can be characterized by combining the knowledge of the distribution of gas in the Galaxy and the observation of gamma rays. We analyze the data from the HAWC Observatory to look for gamma rays in three galactic giant molecular clouds, that are outside the galactic plane ($|b|>5^{\circ}$). We can then test the paradigm that the measured local cosmic-ray flux is the same as the ``sea'' of Galactic cosmic rays. Due to its large field of view, and high duty cycle, HAWC is suitable to search for gamma rays from large structures in the TeV gamma-ray regime. We present here preliminary results from measurements of the Aquila Rift, Hercules and Taurus molecular clouds. }

\FullConference{36th International Cosmic Ray Conference -ICRC2019-\\
		July 24th - August 1st, 2019\\
		Madison, WI, U.S.A.}

\begin{document}

\section{Introduction}\label{sec:info}

Since the discovery of the cosmic rays (CR), we have wondered about their origin and properties.
Direct measurements of CRs measure only the flux and spectrum in the vicinity of the solar system \cite{ams}. It is generally assumed that this spectrum is representative of the CR flux across the galaxy and it is usually referred to as the ``sea'' of cosmic rays. 
This assumption comes from the fact that CRs diffuse through the galaxy due to the ability of the interstellar magnetic field to alter the path of charged particles. If this process occurs on a timescale of millions of years, the distribution of CRs becomes homogeneous and isotropic \cite{crProp}. 

CRs interact with the interstellar matter and radiation fields. This can produce high-energy gamma rays which if observed, give us the ability to measure the propagation and distribution of CRs~\cite{crtogammas, Aharonian2001, barometers}. 

The flux of CRs $\Phi_{CR}$ in a giant molecular cloud (GMC) is related to the flux of gamma rays $F_{\gamma}$, the distance square to the GMC and to the total mass of the GMC as follows:

\begin{equation}\label{eq:modelGCR}
\Phi_{CR} \propto F_{\gamma} \frac{d^2}{M}.
\end{equation}

It is then important to have known distances and masses of several GMCs. 
Eq. \ref{eq:modelGCR} has already being exploited in analyses done using data from the \textit{Fermi}-LAT telescope. In  \cite{fermiCR}, gamma-ray spectra above 300 MeV were used to extract the CR spectra from eight massive clouds. 
Ref. \cite{fermiCR} showed that the derived spectral indices and absolute fluxes of CR protons in the energy interval 10-100 GeV agree with the direct measurements of local CRs by the PAMELA experiment.

A similar study was published by \cite{seaCRFermi}. With their observations, they also observed that the flux of CRs at distances from 0.6 kpc to 12.5 kpc also agrees with the locally measured CR flux measured by AMS \cite{ams}.

\section{Data and Methods}\label{sec:analysis}

The HAWC Observatory is a gamma-ray detector built in Sierra Negra in the Mexican state of Puebla at an altitude of 4100\,m above sea level. It consists of 300 water Cherenkov detectors, each with 4 photomultipliers (PMTs). The PMTs detect the Cherenkov light produced by the secondary particles of extensive showers that cross the water tanks. More information on the HAWC observatory and the way air shower event data are reconstructed is presented in \cite{hawcCrab}.
For the proceedings presented here, we use a dataset that started on 11/2014 and ended on 04/2018 with a livetime of $\sim$1127 days. 

As it was done in previous analyses, we divide the dataset in 9 bins and apply gamma-hadron cuts.
These bins are defined by the ratio of the number of PMTs that participated in the reconstruction of the air-shower event to the total number of active PMTs. 
We will refer to them as fractional bins $f_{hit}$. This is a proxy to an energy variable, where a lower $f_{hit}$ bin corresponds to a lower energy gamma-ray. 
The definition of the bins is presented in Table 2 in \cite{hawcCrab}.

The analysis is performed using the Multi-Mission Maximum Likelihood (3ML) \cite{threeml} framework together with the HAWC Accelerated Likelihood (HAL) framework. 

We compute quasi-differential limits in a similar way as previous HAWC analyses \cite{hawcFB, hawcDM}. 
We define four half-decade energy bins, going from 1 TeV up to 100 TeV. The range of the bins and the midpoint of the range are shown in table \ref{table:energybins}.

\begin{table}[!htp]
    \centering
    \begin{tabular}{c|c|c|c}
    \hline
        Energy Bin  & Lower End [TeV] & Higher End [TeV] & Midpoint [TeV] \\
    \hline
    \hline
         1 &  1 & 3.16 & 1.77\\
         2 & 3.16 &  10.0 & 5.59\\
         3 & 10.0 & 31.6 & 17.7\\
         4 & 31.6 & 100.0 & 55.97\\
    \hline
    \end{tabular}
    \caption{Energy bins definitions used in this analysis.} 
    \label{table:energybins}
\end{table}

In each energy bin, we first maximize the test statistic, based on the likelihood-ratio method. This is defined as
\begin{equation}
TS = 2 \frac{\mathcal{L}(S(\hat{K})+B)}{\mathcal{L}(B)},
\end{equation}
where S is the signal from the alternative hypothesis, while B is the null hypothesis. For the alternative hypothesis, we build a spatial template, based on the Planck survey (see Section \ref{sec:template} below), together with a simple power law as the spectral shape:

\begin{equation}
F_{\gamma}(E)=\frac{dN}{dE} = K \left(\frac{E}{E_{0}}\right)^{-\alpha} 
\end{equation}

where $\alpha$ is the spectral index, which we fixed to be 2.75, $E_0$ is the pivot energy (which in this case is the midpoint specified in table \ref{table:energybins}), and $K$ is the normalization factor, making $\hat{K}$ the normalization that maximizes the $TS$. 

The maximum likelihood estimator is then used as an input to a Markov-Chain Monte Carlo (MCMC) to estimate a distribution of the likelihood around the maximum. In the case of a non-significant detection (i.e. TS${<}25$), we calculate the 95\% credible interval from the estimated likelihood distributions. For the MCMC procedure we assume a uniform prior for the normalization factor in the range of [$0 - 10^{-10}$] TeV$^{-1}$ cm$^{-2}$ s$^{-1}$.

\subsection{Spatial Templates}\label{sec:template}
\begin{figure}[!htp]
\centering
\includegraphics[scale=0.49]{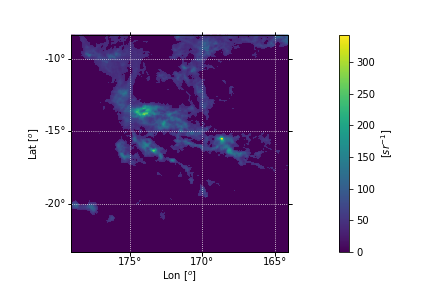}
\includegraphics[scale=0.49]{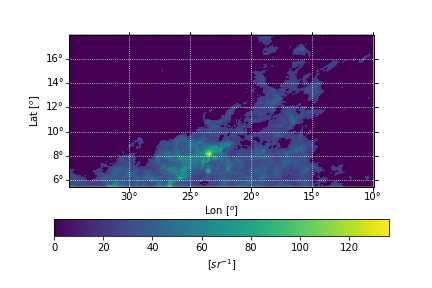}
\includegraphics[scale=0.49]{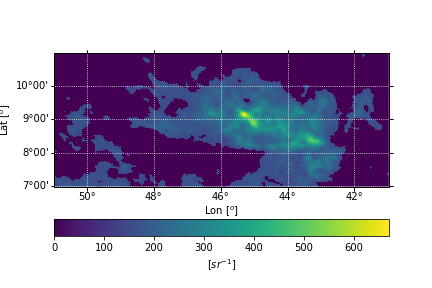}
\caption{GMCs obtained using data from the Planck survey~\cite{planck}. The clouds are Taurus (top-left corner), Aquila (top-right corner) and Hercules (bottom). The figures were obtained after applying the method mentioned in Sec. \ref{sec:template}. The coordinate system shown is galactic coordinates and the units are after normalizing the map.}
\label{fig:clouds}
\end{figure}

The templates were built using data from the Planck survey~\cite{planck} using a similar procedure as in Section 2 of \cite{fermiCR}. 
First we calculate the column density of the cloud from the dust optical depth map at 353~GHz from Eq. \ref{eq:planckCD}. 
\begin{equation}\label{eq:planckCD}
N_{H_2} = \tau_D / \left(\frac{\tau_D}{N_{H_2}}\right)^{ref}
\end{equation}
where the reference value used is $(\tau_D/N_{H_2})^{ref} = 1.18\times10^{-26}$cm$^2$ for 353 GHz \cite{planck}. $\tau_D$ is the dust opacity.
We cut on the opacity value to select the high density regions. For Taurus and Aquila, we use $5\times10^{-5}$ while for Hercules we use  $2.5\times10^{-5}$. Then we normalize the map to the mean column density of the cloud times the size of the spatial bin ($1/(\left< N_{H_2} \right> d\Omega)$), due to the way HAL deals with extended sources. Figure \ref{fig:clouds} shows the three GMCs that we use in this analysis in galactic coordinates. The units are after the normalization process.

Since we will need the mass of the cloud later for calculating the expectation of gamma rays, this is calculated as
\begin{equation}
M_{dust} = N_{H_2} A_{cloud} m_{H_2}
\end{equation}
The area of the cloud $A_{cloud}$ can be expressed as $A_{angular} d^2$, where $d$ is the distance to the cloud, and $m_{H_2}$ is the mass of molecular hydrogen.

\section{Results and Discussion}

We present preliminary results of the analysis described in section \ref{sec:analysis} on the giant molecular clouds Aquila Rift, Hercules, and Taurus. To compare our results with an expectation of gamma rays, we show two models obtained by convolving the CR spectrum measured by AMS\cite{ams} and the pion-to-gamma cross section parametrization from \cite{crossSec}. Spectral information from the molecular cloud can be obtained in two possible ways. One is by using the column density information, which can also be obtained by the Planck survey and then multiplying that by the angular size of the cloud. The second way is by calculating the total mass of the cloud and using the distance to the cloud. This is written as

\begin{equation}\label{eq:m1}
    F_{\gamma} (E_{\gamma} )_1 = N_{H_2} \Omega_{cloud} \int_{T_{p_{min}}}^{T_{p_{max}}} \frac{d\sigma}{dE_{\gamma}}(T_p,E_{\gamma})J(T_p)dT_p,
\end{equation}
and
\begin{equation}\label{eq:m2}
    F_{\gamma} (E_{\gamma} )_2 = 1.25\times10^9 A \int_{T_{p_{min}}}^{T_{p_{max}}} \frac{d\sigma}{dE_{\gamma}}(T_p,E_{\gamma})J(T_p)dT_p.
\end{equation}
$N_{H_2}$ and $\Omega_{cloud}$ are the column density and angular size of the cloud. $A=M_{5}/D^2_{kpc}$; $M_5 = M/10^5 M_{\odot}$; $D_{kpc} = D / 1$kpc are the mass and distance of the cloud. Eq. \ref{eq:m2} can be found in \cite{seaCRFermi}.

Table \ref{table:clouds} shows position, distance, and mass information of the GMCs. 

\begin{table}[]
    \centering
    \begin{tabular}{c|c|c|c|c|c|c|c}
        GMC & Mass[$10^5\, M_\odot$] & Distance [pc] & l[$^{\circ}$] & b[$^{\circ}$] & Dec. [$^{\circ}$] & RA [$^{\circ}$] & Size [sr]\\
        \hline
        \hline
        Taurus & 0.33 & 140$\pm$30 & 171.6 & -15.8  & 30 & 65 & 0.016\\ 
        Aquila Rift & 1.50 & 225$\pm$55 & 25 & 6 & -2.5 & 270 & 0.029 \\ 
        Hercules & 0.090 & 200$\pm$30 & 44 & 9 & 15 & 280 & 0.004\\
        \hline
    \end{tabular}
    \caption{Properties of the GMCs. Distances obtained from \cite{distances} }
    \label{table:clouds}
\end{table}

\begin{figure}[!htp]
\centering
\includegraphics[scale=0.3]{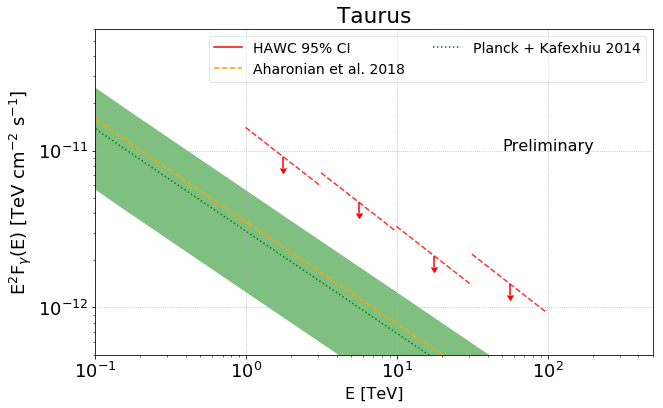}
\includegraphics[scale=0.3]{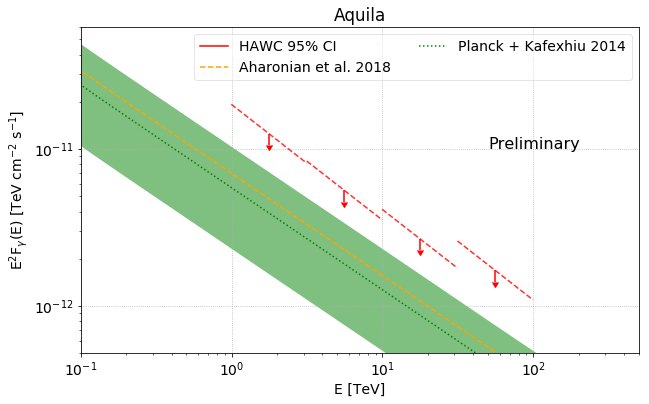}
\includegraphics[scale=0.3]{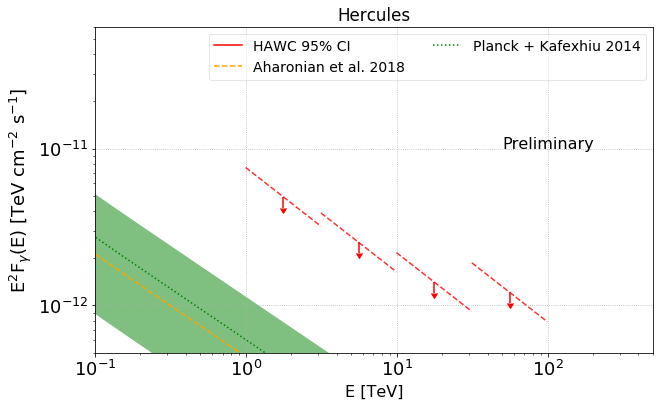}
\caption{95\% C.L. upper limits in the gamma-ray flux. Band represents the uncertainty in the U.L. The results are shown for both of the surveys used. Order: Aquila, Taurus and Hercules. }
\label{fig:uplims}
\end{figure}

Figure \ref{fig:uplims} shows the upper limits together with the expected model in green for Eq. \ref{eq:m1} and orange for Eq. \ref{eq:m2}. The green band corresponds to the range from the 10th to 90th percentiles of the column density distribution, while the line corresponds to the median. 
The credible interval upper limits can be found in Table \ref{table:credint}.

\begin{table}
    \centering
    \begin{tabular}{c|c|c|c|c}
    \hline
    GMC & 1 - 3.16 TeV & 3.16 - 10.0 TeV & 10.0 - 31.6 TeV & >31.6 TeV \\
        & [$\times10^{-12}$] & [$\times10^{-13}$] & [$\times10^{-15}$] & [$\times10^{-16}$] \\
    \hline
    \hline
     Taurus & 2.9 &  1.5 & 6.8 & 4.5  \\
     Aquila & 4.0 &  1.7 & 8.6 & 5.3 \\
     Hercules & 1.6 & 0.8 & 4.5 & 3.8\\
     \hline
    \end{tabular}
    \caption{Quasi-Differential 95\% Credible Intervals  in TeV$^{-1}$ cm$^{-2}$ s$^{-1}$}
    \label{table:credint}
\end{table}

\section{Conclusion}
With the purpose of increasing the measurements of gamma rays and cosmic rays from giant molecular clouds, we searched for gamma rays from three molecular clouds using data from the HAWC observatory. Since no significant excess was observed, we calculated upper limits at the 95$\%$ C.L. We see that the expected gamma ray flux from pure hadronic interactions of the cosmic ray flux with passive molecular clouds is below $10^{-11}$ TeV$^{-1}$ cm$^{-2}$ s$^{-1}$ above 10\,TeV. The upper limits are above the expected range of values from $pp$ interactions(See Fig.\ref{fig:uplims}). 

\acknowledgments
\small{
We acknowledge the support from: the US National Science Foundation (NSF) the US Department of Energy Office of High-Energy Physics; 
the Laboratory Directed Research and Development (LDRD) program of Los Alamos National Laboratory; 
Consejo Nacional de Ciencia y Tecnolog\'{\i}a (CONACyT), M{\'e}xico (grants 271051, 232656, 260378, 179588, 239762, 254964, 271737, 258865, 243290, 132197, 281653)(C{\'a}tedras 873, 1563, 341), Laboratorio Nacional HAWC de rayos gamma; 
L'OREAL Fellowship for Women in Science 2014; 
Red HAWC, M{\'e}xico; 
DGAPA-UNAM (grants AG100317, IN111315, IN111716-3, IA102715, IN109916, IA102019, IN112218); 
VIEP-BUAP; 
PIFI 2012, 2013, PROFOCIE 2014, 2015; 
the University of Wisconsin Alumni Research Foundation; 
the Institute of Geophysics, Planetary Physics, and Signatures at Los Alamos National Laboratory; 
Polish Science Centre grant DEC-2014/13/B/ST9/945, DEC-2017/27/B/ST9/02272; 
Coordinaci{\'o}n de la Investigaci{\'o}n Cient\'{\i}fica de la Universidad Michoacana; Royal Society - Newton Advanced Fellowship 180385. Thanks to Scott Delay, Luciano D\'{\i}az and Eduardo Murrieta for technical support.
}

\bibliographystyle{ICRC}
\bibliography{references}

%

\end{document}